\documentclass[aps,prc,twocolumn,superscriptaddress,floatfix,amsmath,amssymb,nofootinbib,longbibliography]{revtex4-2}
\usepackage[utf8]{inputenc}
\usepackage[T1]{fontenc}

\usepackage{xcolor}
\usepackage{graphicx}
\usepackage{microtype}

\usepackage{braket}

\usepackage[colorlinks=true,citecolor=blue,linkcolor=blue,urlcolor=blue]{hyperref}

\newcommand{\imsrg}{\mbox{IMSRG}}
\newcommand{\imsrgthreenseven}{\mbox{IMSRG(3)-$N^7$}}
\newcommand{\imsrgthree}{\mbox{IMSRG(3)}}
\newcommand{\imsrgfour}{\mbox{IMSRG(4)}}
\newcommand{\imsrgtwo}{\mbox{IMSRG(2)}}

\newcommand{\vsimsrg}{\mbox{VS-IMSRG}}
\newcommand{\vsimsrgthreenseven}{\mbox{VS-IMSRG(3)-$N^7$}}
\newcommand{\vsimsrgthree}{\mbox{VS-IMSRG(3)}}
\newcommand{\vsimsrgtwo}{\mbox{VS-IMSRG(2)}}
\newcommand{\lccsdt}{\mbox{$\Lambda$-CCSD(T)}}
\newcommand{\ccsdtone}{\mbox{CCSDT-1}}

\newcommand{\magic}{1.8/2.0~(EM)}

\newcommand{\MeV}{\text{MeV}}

\newcommand{\fm}{\text{fm}}
\newcommand{\elem}[2]{\ensuremath{^{\text{#2}}\text{#1}}}

\newcommand{\nn}{\ensuremath{\text{NN}}}
\newcommand{\nnn}{\ensuremath{\text{3N}}}

\newcommand{\vnn}{\ensuremath{V_\nn}}
\newcommand{\vnnn}{\ensuremath{V_\nnn}}

\newcommand{\nuhamil}{\textsc{NuHamil}}

\newcommand{\rch}{\ensuremath{R_\text{ch}}}

\newcommand{\rp}{\ensuremath{R_\text{p}}}
\newcommand{\rptwo}{\ensuremath{R_\text{p}^2}}
\newcommand{\rn}{\ensuremath{R_\text{n}}}
\newcommand{\rntwo}{\ensuremath{R_\text{n}^2}}
\newcommand{\rsotwo}{\ensuremath{r_\text{so}^2}}
\newcommand{\rprottwo}{\ensuremath{r_\text{p}^2}}
\newcommand{\rneuttwo}{\ensuremath{r_\text{n}^2}}
\newcommand{\ecorr}{\ensuremath{E_\text{corr}}}
\newcommand{\rskin}{\ensuremath{R_\text{skin}}}

\newcommand{\hw}{\ensuremath{\hbar\omega}}
\newcommand{\ho}{\ensuremath{\text{HO}}}
\newcommand{\hf}{\ensuremath{\text{HF}}}
\renewcommand{\nat}{\ensuremath{\text{NAT}}}
\newcommand{\emax}{\ensuremath{e_\text{max}}}
\newcommand{\emaxmeanfield}{\ensuremath{e_\text{max}^\ho}}
\newcommand{\ethreemaxmeanfield}{\ensuremath{E_\text{3max}^\ho}}
\newcommand{\emaxthreeb}{\ensuremath{e_\text{max,3b}}}
\newcommand{\ethreemax}{\ensuremath{E_\text{3max}}}

\newcommand{\crea}[1]{\ensuremath{a^{\dagger}_{#1}}}

\newcommand{\comm}[2]{\ensuremath{\left[#1,#2 \right]}}

\begin{document}

\title{Improved structure of calcium isotopes from ab initio calculations}

\author{M.~Heinz}
\email{heinzmc@ornl.gov}
\affiliation{Technische Universit\"at Darmstadt, Department of Physics, 64289 Darmstadt, Germany}
\affiliation{ExtreMe Matter Institute EMMI, GSI Helmholtzzentrum f\"ur Schwerionenforschung GmbH, 64291 Darmstadt, Germany}
\affiliation{Max-Planck-Institut f\"ur Kernphysik, Saupfercheckweg 1, 69117 Heidelberg, Germany}
\affiliation{National Center for Computational Sciences, Oak Ridge National Laboratory, Oak Ridge, TN 37831, USA}
\affiliation{Physics Division, Oak Ridge National Laboratory, Oak Ridge, TN 37831, USA}

\author{T.~Miyagi}
\affiliation{Center for Computational Sciences, University of Tsukuba, 1-1-1 Tennodai, Tsukuba 305-8577, Japan}
\affiliation{Technische Universit\"at Darmstadt, Department of Physics, 64289 Darmstadt, Germany}
\affiliation{ExtreMe Matter Institute EMMI, GSI Helmholtzzentrum f\"ur Schwerionenforschung GmbH, 64291 Darmstadt, Germany}
\affiliation{Max-Planck-Institut f\"ur Kernphysik, Saupfercheckweg 1, 69117 Heidelberg, Germany}

\author{S.~R.~Stroberg}
\affiliation{Department of Physics and Astronomy, University of Notre Dame, Notre Dame, IN 46556, USA}
\affiliation{Physics Division, Argonne National Laboratory, Lemont, IL 60439, USA}

\author{A.~Tichai}
\affiliation{Technische Universit\"at Darmstadt, Department of Physics, 64289 Darmstadt, Germany}
\affiliation{ExtreMe Matter Institute EMMI, GSI Helmholtzzentrum f\"ur Schwerionenforschung GmbH, 64291 Darmstadt, Germany}
\affiliation{Max-Planck-Institut f\"ur Kernphysik, Saupfercheckweg 1, 69117 Heidelberg, Germany}

\author{K.~Hebeler}
\affiliation{Technische Universit\"at Darmstadt, Department of Physics, 64289 Darmstadt, Germany}
\affiliation{ExtreMe Matter Institute EMMI, GSI Helmholtzzentrum f\"ur Schwerionenforschung GmbH, 64291 Darmstadt, Germany}
\affiliation{Max-Planck-Institut f\"ur Kernphysik, Saupfercheckweg 1, 69117 Heidelberg, Germany}

\author{A.~Schwenk}
\affiliation{Technische Universit\"at Darmstadt, Department of Physics, 64289 Darmstadt, Germany}
\affiliation{ExtreMe Matter Institute EMMI, GSI Helmholtzzentrum f\"ur Schwerionenforschung GmbH, 64291 Darmstadt, Germany}
\affiliation{Max-Planck-Institut f\"ur Kernphysik, Saupfercheckweg 1, 69117 Heidelberg, Germany}

\begin{abstract}

The in-medium similarity renormalization group (\imsrg{}) is a powerful and flexible many-body method to compute the structure of nuclei starting from nuclear forces.
Recent developments have extended the \imsrg{} from its standard truncation at the normal-ordered two-body level, the \imsrgtwo{}, to a precision approximation including normal-ordered three-body operators, the \imsrgthreenseven{}.
This improvement provides a more precise solution to the many-body problem and makes it possible to quantify many-body uncertainties in \imsrg{} calculations.
We explore the structure of \elem{Ca}{44,48,52} using the \imsrgthreenseven{}, focusing on understanding existing discrepancies of the \imsrgtwo{} to experimental results.
We find a significantly better description of the first $2^+$ excitation energy of \elem{Ca}{48}, improving the description of the shell closure at $N=28$.
At the same time, we find that the \imsrgthreenseven{} corrections to charge radii do not resolve the systematic underprediction of the puzzling large charge radius difference between \elem{Ca}{52} and \elem{Ca}{48}.
We present estimates of many-body uncertainties of \imsrgtwo{} calculations applicable also to other systems based on the size extensivity of the method.

\end{abstract}

\maketitle

\section{Introduction}
\label{sec:intro}

The nuclear structure of calcium isotopes has long been studied, both experimentally and theoretically, but there are still many open questions about their structure, especially in neutron-rich systems.
New neutron-rich magic numbers at $N=32$ and $N=34$ are suggested by some experiments~\cite{Gallant2012PRL_Ca52, Wienholtz2013N_CaMassesNuclearForces, Steppenbeck2013N_Ca54, Enciu2022PRL_Ca52ExtendedOrbitals}
but brought into question by others (also in neighboring elements)~\cite{GarciaRuiz2016NP_CaRadii,Koszorus2021NP_PotassiumRadiiN28,Liu2019PRL_Ar52}.
Moreover, \elem{Ca}{60} has been observed~\cite{Tarasov2018PRL_Ca60}, but little is known about its structure~\cite{Cortes2020PLB_Ti62, Chen2023PLB_Ca58LevelStructures}, which will have important implications for the neutron drip line in calcium.

Theoretically, the calcium isotopes have been extensively studied using both phenomenological and ab initio approaches.
For ab initio nuclear structure theory, the description of medium-mass nuclei is made possible through the use of approximate, systematically improvable many-body methods~\cite{Dickhoff2004PPNP_SCGFReview, Lee2009PPNP_LatticeReview, Hagen2014RPP_CCReview, Hergert2016PR_IMSRG, Tichai2020FP_MBPTReview, Hergert2020FP_AbInitioReview} with computational scaling in mass number mild enough to allow for the description of systems as heavy as \elem{Pb}{208}~\cite{Morris2018PRL_Sn100, Miyagi2022PRC_NO2B, Hu2022NP_Pb208, Hebeler2023PRC_JacobiNO2B, Arthuis2024arxiv_LowResForces}.
Ab initio studies have been successful in predicting the trends of ground-state energies, two-neutron separation energies, excitation spectra, and neutron skins of calcium isotopes~\cite{Gallant2012PRL_Ca52, Wienholtz2013N_CaMassesNuclearForces,GarciaRuiz2016NP_CaRadii,Hagen2012PRL_CaEOMCC, Soma2014PRC_SCGFChains, Hergert2014PRC_MRIMSRG, Hagen2016PRL_Ni78, Hagen2016NP_Ca48Skin, Stroberg2017PRL_VSIMSRG, Simonis2017PRC_ChiralSaturationNuclei, Tichai2018PLB_BMBPT, Stroberg2021PRL_AbInitioLimits, Tichai2024PLB_BCC} but struggle to explain the trends in charge radii~\cite{GarciaRuiz2016NP_CaRadii, Miyagi2020PRC_MultiShellIMSRG}.
For all of these studies, the many-body methods employed are approximate, but the uncertainty due to the many-body approximation is not systematically explored, opening the question of whether existing discrepancies are due to higher-order many-body physics not captured by the methods used.
Notably, for the $2^+$ energy of \elem{Ca}{48} it was shown in coupled-cluster theory, one such many-body approach, that extending the method to higher orders yielded important corrections improving the agreement with experiment~\cite{Hagen2016PRL_Ni78}.

In this work, we revisit the ab initio description of the structure of calcium isotopes using the in-medium similarity renormalization group (\imsrg{})~\cite{Tsukiyama2011PRL_IMSRG,Tsukiyama2012PRC,Hergert2016PR_IMSRG,Stroberg2017PRL_VSIMSRG,Stroberg2019ARNPS_VSIMSRG}.
The \imsrg{} is typically approximated by truncating all operators at the normal-ordered two-body level, the \imsrgtwo{}, but recent developments have relaxed this approximation to also include normal-ordered three-body operators, the \imsrgthreenseven{}~\cite{Heinz2021PRC_IMSRG3}.
This makes the method more precise and gives insight into the many-body uncertainties of the \imsrgtwo{}~\cite{Heinz2021PRC_IMSRG3, Stroberg2024PRC_IMSRG3}.
In this work, we use this improved precision to investigate the structure of \elem{Ca}{44,48,52} to understand existing discrepancies with experiment.

This paper is structured as follows: In Sec.~\ref{sec:Method}, we introduce our theoretical approach.
In Sec.~\ref{sec:Results}, we compute the structure of \elem{Ca}{44}, \elem{Ca}{48}, and \elem{Ca}{52}.
We first perform a systematic investigation of the improved structure of \elem{Ca}{48}.
We then turn our attention to the charge radius trends in calcium isotopes.
We also investigate the improvements to predicted excitation spectra.
Our systematic study allows us to provide some general estimates of \imsrgtwo{} uncertainties that will be applicable to other studies.
Finally, we conclude in Sec.~\ref{sec:Conclusion}.

\section{Method}
\label{sec:Method}

In this work, we solve the many-body Schrödinger equation for the intrinsic nuclear Hamiltonian of a nucleus with mass number $A$
\begin{equation}
    H = T_\text{int} + \vnn{} + \vnnn{}\,,
\end{equation}
with the intrinsic kinetic energy (with the center-of-mass contribution removed) $T_\text{int}$ and a given set of NN and 3N potentials \vnn{} and \vnnn{}~\cite{Hergert2016PR_IMSRG}.
We use the IMSRG, a standard method for ab initio nuclear structure calculations in medium-mass and heavy nuclei~\cite{Tsukiyama2011PRL_IMSRG, Hergert2016PR_IMSRG, Stroberg2017PRL_VSIMSRG, Stroberg2019ARNPS_VSIMSRG, Heinz2021PRC_IMSRG3}.

\subsection{Computational basis and reference state}
\label{sec:Basis}

We start by constructing our computational single-particle basis with states
\begin{equation}
    \label{eq:SingleParticleBasis}
    \ket{p} = \ket{n l j m_j m_t} = \crea{p} \ket{0}.
\end{equation}
Here $p$ is a collective index for the quantum numbers of the state:
the principal quantum number $n$, the orbital angular momentum $l$, the total angular momentum $j$ (from coupling $l$ with spin $s=1/2$ for nucleons), its projection $m_j$, and the isospin projection $m_t$ distinguishing protons and neutrons.
The harmonic oscillator (\ho) ``energy'' of a state is $e_p = 2n + l$, and our computational basis includes states with $e_p \leq \emax$.

We expand all states and all operators in the eigenbasis of an isotropic harmonic oscillator with $\hw = 16~\MeV$, including \ho{} states $\ket{p}_\ho$ with $e_p^\ho \leq \emaxmeanfield = 16$.
For \nnn{} potentials, we employ an additional truncation in the three-body basis $\ket{pqr}_\ho$, including only states with $e_p^\ho + e_q^\ho + e_r^\ho \leq \ethreemaxmeanfield = 24$~\cite{Miyagi2022PRC_NO2B}.
We generated these matrix elements using the \nuhamil{} code~\cite{Miyagi2023EPJA_NuHamil}.

In our computational basis, we construct the reference state for our system of interest,
\begin{equation}
    \ket{\Phi} = \prod_{i = 1}^{A} \crea{p_i} \ket{0},
\end{equation}
from the $A$ energetically lowest states employing ensemble normal ordering where necessary~\cite{Stroberg2017PRL_VSIMSRG}.
We use a Hartree-Fock (\hf) basis for the occupied states and a natural orbital (\nat) basis orthogonalized with respect to the occupied \hf{} states for the remaining unoccupied states.
This construction detailed in Appendix~\ref{app:HFNAT} combines the energetically optimal \hf{} reference state with the improved model-space convergence of the \nat{} basis~\cite{Tichai2019PRC_NAT,Novario2020PRC_DeformedCC,Hoppe2021PRC_NAT}.

Given our computational basis and reference state, we normal order all operators with respect to the reference state.
For the Hamiltonian, we get
\begin{align}
    H & = E + f + \Gamma + W\,,\label{eq:FullHamiltonian}
\end{align}
with the normal-ordered zero- through three-body parts $E$, $f$, $\Gamma$, and $W$~\cite{Hergert2016PR_IMSRG}.
Here $E$ is simply the reference-state expectation value, the \hf{} energy.
In this work, we discard the residual three-body Hamiltonian $W$ in Eq.~\eqref{eq:FullHamiltonian} at this stage, employing the well-established normal-ordered two-body (NO2B) approximation~\cite{Hagen2007PRC_CC3N,Roth2012PRL_NO2B3N,Binder2014PLB_NO2BCC, Djarv2021PRC_NO2BApprox},
\begin{equation}
    \label{eq:NO2B}
    H = E + f + \Gamma\,.
\end{equation}

\subsection{In-medium similarity renormalization group}
\label{sec:IMSRG}

The IMSRG generates a continuous, tailored unitary transformation of the Hamiltonian
\begin{equation}
    H(s) = U(s)\, H\, U^{\dagger}(s)
\end{equation}
via the solution of the IMSRG flow equation
\begin{equation}
    \label{eq:IMSRGFlow}
    \frac{dH(s)}{ds} = \comm{\eta(s)}{H(s)},
\end{equation}
integrating the flow parameter $s$ from $s=0$ to $s\rightarrow \infty$.
The unitary transformation is determined by the choice of the generator $\eta$.
Two common approaches are
the single-reference IMSRG~\cite{Tsukiyama2011PRL_IMSRG,Hergert2016PR_IMSRG}, where the reference state is decoupled from its excitations in the transformed Hamiltonian, directly giving the ground-state energy $E(s\rightarrow\infty)$ and wave function;
and the valence-space IMSRG (VS-IMSRG)~\cite{Tsukiyama2012PRC,Stroberg2017PRL_VSIMSRG,Stroberg2019ARNPS_VSIMSRG}, where a core and a valence space are decoupled from the remaining states in the transformed Hamiltonian and a final diagonalization of the valence-space Hamiltonian via shell-model techniques gives the ground-state energy and wave function.
Other ground-state properties can be computed by applying the same unitary transformation to the operator of interest,
\begin{equation}
    O(s) = U(s)\, O\, U^{\dagger}(s)\,,
\end{equation}
evaluating its expectation value in the IMSRG ground state.
We use the Magnus formulation of the IMSRG equations above~\cite{Morris2015PRC_Magnus}, giving direct access to the unitary transformation $U(s) = e^{\Omega(s)}$ in terms of the Magnus operator $\Omega(s)$.

IMSRG calculations are typically truncated at the normal-ordered two-body level, the IMSRG(2), keeping up to normal-ordered two-body terms for the Hamiltonian, the generator, and all operators, e.g.,
\begin{equation}
    H(s) = E(s) + f(s) + \Gamma(s)\,.
\end{equation}
This is an approximation, as the commutator in Eq.~\eqref{eq:IMSRGFlow} will induce normal-ordered three-body and also higher-body contributions if not truncated.
This approximation can be relaxed by also including normal-ordered three-body operators, yielding the IMSRG(3)~\cite{Hergert2016PR_IMSRG, Heinz2021PRC_IMSRG3, Stroberg2024PRC_IMSRG3}.

We explore IMSRG(3) calculations using the \imsrgthreenseven{} truncation~\cite{Heinz2021PRC_IMSRG3}, which is the same as the IMSRG(3N7) truncation of Ref.~\cite{Stroberg2024PRC_IMSRG3}.
In this truncation, all terms in the IMSRG equations that scale as $O(N^7)$ or milder in the size of the single-particle basis are included, and three-body operators are included fully nonperturbatively in the calculations:
\begin{equation}
    H(s) = E(s) + f(s) + \Gamma(s) + W(s)\,.
\end{equation}
This is to be contrasted with the IMSRG(2*) of Refs.~\cite{Morris2016_Thesis,Stroberg2024PRC_IMSRG3} and the IMSRG(3f$_2$) of Ref.~\cite{He2024PRC_IMSRG3f2}, which include three-body corrections to the IMSRG(2) in ways that do not include explicit $s$-dependent three-body operators $W(s)$.
Given the fact that we start from an NO2B-truncated Hamiltonian [see Eq.~\eqref{eq:NO2B}], the three-body part $W(s)$ captures induced effective three-body interactions, which in turn also modify $E(s)$, $f(s)$, and $\Gamma(s)$ to perform a more accurate unitary transformation and provide a more precise result for the ground-state energy and low-lying spectrum.

Three-body operators are exceptionally challenging to treat computationally, in large part due to the immense cost of storing all their matrix elements $\braket{pqr | W |stu}$.
For this reason, we further restrict the basis we use for three-body operators beyond the level of our single-particle basis [defined in Eq.~\eqref{eq:SingleParticleBasis} and truncated based on \emax{}].
We truncate the basis of three-body states $\ket{pqr}$ such that
\begin{align}
    e_p,\:e_q,\:e_r &\leq \emaxthreeb \,,\\
    e_p + e_q + e_r &\leq \ethreemax \,,
\end{align}
and similarly for the $\ket{stu}$ states.
Fully relaxing this truncation means taking $\emaxthreeb \rightarrow \emax$, $\ethreemax \rightarrow 3 \emax$, but in practice this is unreachable and we explore how well our results are converged with respect to the two parameters \emaxthreeb{}, \ethreemax{}.

In \vsimsrg{} calculations, the final valence-space Hamiltonian must be diagonalized using shell-model techniques.
Many shell-model solvers are restricted to one- and two-body interactions,
including the \textsc{kshell} code we use in this work~\cite{Shimizu2019CPC_KSHELL}.
When doing calculations with the \vsimsrgthreenseven{},
there are, however, also three-body valence-space interactions included in the three-body operators.
This raises the question of how one should treat these interactions when the shell-model solver cannot include them in the diagonalization.

In this work, we leverage the well-established cluster hierarchy of many-body interactions in nuclear structure calculations~\cite{Friman2011Book_FLT3N, Stroberg2019ARNPS_VSIMSRG},
which states that one-body interactions are more important than two-body interactions, which in turn are more important than three-body interactions.
Motivated by this, we keep the three-body interactions while we solve the \imsrg{} equations, providing a more precise unitary transformation of the Hamiltonian.
However, once our desired decoupling has been reached, we truncate the three-body interactions because at this point their contribution when solving the remaining valence-space problem is expected to be small.
Essentially, we set $W(s=0) = 0$ to apply the standard NO2B truncation, and similarly we set $W(s\rightarrow\infty) = 0$ to apply a similar truncation before computing our valence-space interactions.
These are then computed using the standard ensemble normal-ordering procedure~\cite{Stroberg2016PRC_VSIMSRGLong, Stroberg2017PRL_VSIMSRG}.

In the \imsrg{}, ground-state and excited-state energies can be simply computed from the transformed Hamiltonian $H(s)$.
In this work, we also consider ground-state radius observables, specifically charge radii \rch{} and neutron skins \rskin{}.
These are computed by consistently unitarily transforming the associated radius operators and evaluating ground-state expectation values.
We compute the charge radius squared as
\begin{equation}
  \label{eq:Rch}
  R^2_\text{ch} = 
  \langle \rptwo \rangle
  + \langle \rsotwo \rangle
  + \rprottwo
  + \frac{N}{Z} \rneuttwo
  + \frac{3}{4M^2}\,,
\end{equation}
based on the point-proton radius squared \rptwo{}, the spin-orbit correction \rsotwo{}~\cite{Ong2010PRC_SpinOrbit}, the proton charge radius squared $\rprottwo = 0.771~\fm^2$, the neutron charge radius squared $\rneuttwo = -0.115~\fm^2$, and the Darwin-Foldy correction using the nucleon mass $M = 938.919~\MeV$~\cite{Friar1997PRA_DarwinFoldy}.\footnote{
    The proton charge radius value has recently been updated~\cite{Workman2022PTEP_PDG2022} with a value $\rprottwo = 0.7071\,(7)~\fm^2$, which reduces the charge radius of the systems we consider by roughly $0.01~\fm$.
    For consistent comparison with results from past studies~\cite{Hagen2016NP_Ca48Skin},
    we employ the outdated value in this work.
}
We provide details on these operators including a correction to a published mistake in the spin-orbit radius operator~\cite{Ong2010PRC_SpinOrbit}, which was pointed out by Martin Hoferichter, in Appendix~\ref{app:RadiusOps}.
Furthermore the neutron skin
\begin{equation}
  \label{eq:Rskin}
  R_\text{skin} = 
  \langle \rntwo \rangle^{1/2} -
  \langle \rptwo \rangle^{1/2}\,
\end{equation}
is computed as the difference of the point-neutron and point-proton charge radii.

\section{Results}
\label{sec:Results}

In the following, we explore the structure of neutron-rich calcium isotopes using the IMSRG.
We perform all calculations with the \textsc{imsrg++} code~\cite{Stroberg2024_IMSRGGit}.
Unless otherwise stated, we use the \magic{} Hamiltonian~\cite{Hebeler2011PRC_SRG3NFits}.
We construct our HF+NAT computational basis using an HO basis consisting of 17 major shells ($\emaxmeanfield = 16$) with three-body matrix elements truncated at $\ethreemaxmeanfield = 24$.
We truncate this basis to $\emax=10$ for all calculations presented here, which is sufficiently converged to investigate the \imsrgthreenseven{} corrections we are interested in.
For example, for \elem{Ca}{48},
the ground-state energy is converged within 500~keV,
the $2^+$ energy is converged within 150~keV,
the charge radius is converged within 0.002~fm,
and the neutron skin is converged within 0.0015~fm.
We provide additional details on our basis and model-space truncation in Appendix~\ref{app:HFNAT}.
We investigate both single-reference and valence-space \imsrgthreenseven{} calculations.
For our valence-space calculations of \elem{Ca}{44}, \elem{Ca}{48}, and \elem{Ca}{52},
we employ a $0\hbar\omega$ valence space with a \elem{Ca}{40} core
and four, eight, and twelve valence neutrons, respectively, interacting in a valence space consisting of the neutron $1f_{7/2}$, $2p_{3/2}$, $2p_{1/2}$, and $1f_{5/2}$ orbitals.

We primarily compare \imsrgthreenseven{} results as a function of \emaxthreeb{} and \ethreemax{} with results from coupled-cluster theory,
where coupled-cluster with singles and doubles (CCSD) is similar to the \imsrgtwo{} in computational cost and perturbative content~\cite{Hagen2014RPP_CCReview,Hergert2016PR_IMSRG,Heinz2021PRC_IMSRG3,Stroberg2024PRC_IMSRG3}.
Like the \imsrgthree{}, coupled-cluster with singles, doubles, and triples (CCSDT) is too computationally expensive~\cite{Noga1987JCP_CCSDT,Scuseria1988CPL_CCSDT}, which has led to the proliferation of many approximate treatments of triples.
The notable methods we compare with are \lccsdt, where the effects of triples are computed perturbatively based on a CCSD calculation~\cite{Taube2008JCP_LambdaCCSDT,Taube2008JCP_LambdaCCSDTDerivatives,Binder2013PRC_LambdaCCSDT3N,Hagen2014RPP_CCReview},
and \ccsdtone, where the effects of leading triples are solved for iteratively~\cite{Lee1984JCP_CCSDT1}.
\lccsdt{} is most similar to recent perturbatively improved IMSRG approximations~\cite{He2024PRC_IMSRG3f2},
and \ccsdtone{} is most similar to the \imsrgthreenseven{}~\cite{Heinz2021PRC_IMSRG3, Stroberg2024PRC_IMSRG3}.

\subsection{Structure of \texorpdfstring{\elem{Ca}{48}}{calcium-48}}

\begin{figure}
    \includegraphics{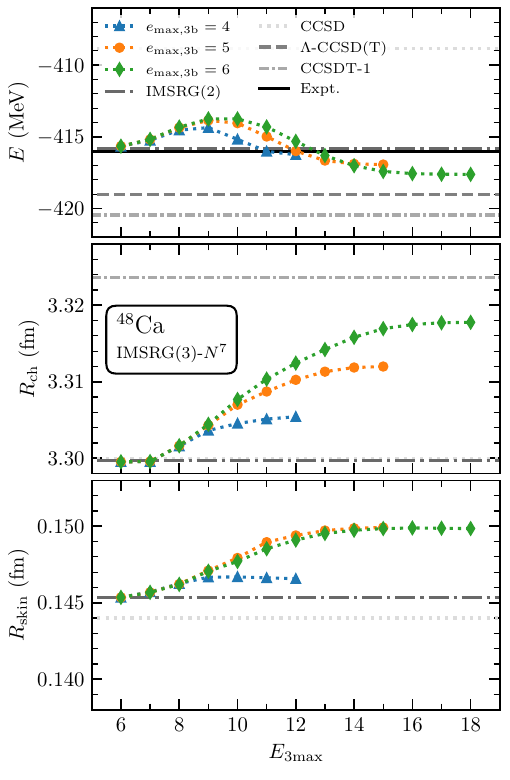}
    \caption{
        \label{fig:IMSRG3GroundState}
        Ground-state energy (top), charge radius (middle), and neutron skin (bottom) of \elem{Ca}{48} computing using the \imsrgthreenseven{}.
        Reference \imsrgtwo{} values (dot-dashed line) are compared with
        \imsrgthreenseven{} predictions for increasing three-body truncations \emaxthreeb{} and \ethreemax{},
        with $\emaxthreeb = 4$ (blue triangles), 5 (orange circles), 6 (green diamonds) and \ethreemax{} ranging from 6 to $3 \emaxthreeb{}$.
        We compare these results with values from coupled-cluster theory~\cite{Hagen2016NP_Ca48Skin,Simonis2019EPJA_CCEMResponses,Bonaiti2024private},
        including CC with singles and doubles (CCSD, dotted),
        CC with singles, doubles, and perturbative triples [$\Lambda$-CCSD(T), dashed],
        and CC with leading iterated triples (CCSDT-1, narrow dot-dashed),
        and the experimental ground-state energy of \elem{Ca}{48}~\cite{Wang2021CPC_AME2020}.
    }
\end{figure}

We start by considering the structure of \elem{Ca}{48}.
In Fig.~\ref{fig:IMSRG3GroundState}, we compute its ground-state properties with the single-reference \imsrg{}.
In the top panel, we show the ground-state energy.
We see that the \imsrgtwo{} predicts an energy very close to the experimental ground-state energy,
reflecting the well-established fact that the \magic{} Hamiltonian accurately reproduces ground-state energies in medium-mass systems.
When comparing to results from coupled-cluster theory, we see that the \imsrgtwo{} is closer to \lccsdt{} and \ccsdtone{} than CCSD.
This behavior has been analyzed using perturbative techniques~\cite{Hergert2016PR_IMSRG,Morris2016_Thesis,Stroberg2024PRC_IMSRG3}, where it was found that the IMSRG(2) undercounts a few fourth-order quadruples contributions relative to CCSD.
These quadruples are generally repulsive for soft Hamiltonians, while triples missing from both CCSD and the \imsrgtwo{} are attractive, and the two contributions largely cancel in fourth-order perturbation theory~\cite{Hergert2016PR_IMSRG, Stroberg2024PRC_IMSRG3}.
Such analyses are complicated by the different nonperturbative content of the two methods.
Nonetheless, the \imsrgtwo{} and \lccsdt{} have been observed to give very similar predictions for ground-state energies in a broad range of applications~\cite{Hagen2016NP_Ca48Skin, Simonis2017PRC_ChiralSaturationNuclei, Hergert2020FP_AbInitioReview, Hu2022NP_Pb208}, including many with harder Hamiltonians, while CCSD misses about 10\,\% of the correlation energy~\cite{Hagen2014RPP_CCReview, Tichai2024PLB_BCC}.

\begin{figure*}
    \includegraphics{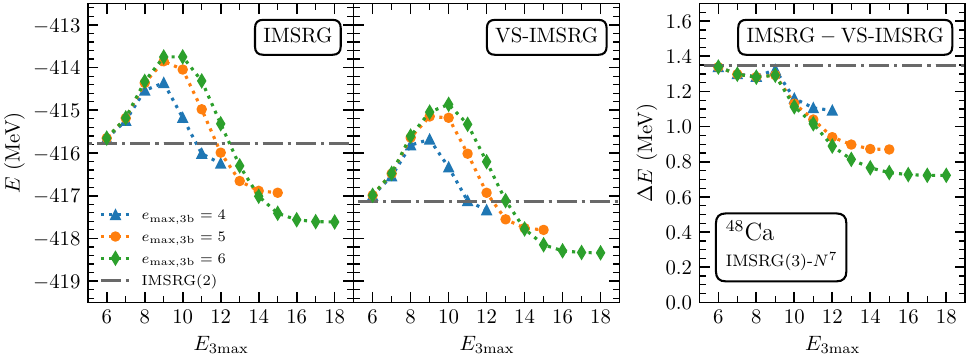}
    \caption{
        \label{fig:VSIMSRGComparisonEnergy}
        The ground-state energy of \elem{Ca}{48} computed by the \imsrg{} and \vsimsrg{}.
        The left and center panels show \imsrgthreenseven{} results for increasing three-body truncations \emaxthreeb{} and \ethreemax{} obtained using the single-reference and valence-space \imsrg{} formulations, respectively, with the difference between the two results ($\Delta E = E_\text{\imsrg} - E_\text{\vsimsrg}$) displayed in the right panel.
    }
\end{figure*}

Comparing the \imsrgthreenseven{} and \imsrgtwo{} predictions for the ground-state energy, we find overall very small corrections.
For increasing model-space parameters \emaxthreeb{} and \ethreemax{},
the \imsrgthreenseven{} energy smoothly converges toward around $-418~\MeV{}$, roughly 2~MeV lower than the \imsrgtwo{} result.
The remaining model-space uncertainty is around 1.5~\MeV{} based on the difference from $\emaxthreeb{} = 5$ and $\emaxthreeb{} = 6$,
and the converged result is likely close to the \lccsdt{} result.
Overall, the \imsrgthree{} corrections are on the order of 2\,\% of the correlation energy, which for \elem{Ca}{48} for the \magic{} Hamiltonian is about 110~MeV.

For the charge radius of \elem{Ca}{48}, the \imsrgtwo{} and CCSD are in excellent agreement.
The \imsrgthreenseven{} corrections produce slightly larger radii.
Here, the convergence in \emaxthreeb{} is slower than for the ground-state energy, and as a result the remaining model-space uncertainty is larger.
Nonetheless, we see that the \imsrgthreenseven{} corrections are consistent in sign and magnitude with the \ccsdtone{} corrections,
and the fully converged result likely lies somewhere between 3.33--3.34~fm.\footnote{
    Extrapolating the current trends based on the similar convergence pattern of IMSRG(2) charge radii in \emax{} (see Fig.~\ref{fig:Ca48IMSRG2}) yields a charge radius of 3.335~fm.
    To be conservative, we give the range above, because it is not guaranteed that there are no large contributions to the charge radius for $\emaxthreeb > 6$.
}
These corrections of 1\,\% to the charge radius are also generally small, especially compared to the uncertainties of the input Hamiltonians.

In the bottom panel of Fig.~\ref{fig:IMSRG3GroundState}, we consider the neutron skin of \elem{Ca}{48}, $R_\text{skin} = \rn - \rp$.
Differential quantities like the neutron skin have the useful quality that many systematic theory uncertainties, both in the Hamiltonian and the many-body calculation, are correlated in the predictions of the point-proton and point-neutron radii and thus largely cancel in the difference~\cite{Hagen2016NP_Ca48Skin,Arthuis2024arxiv_LowResForces}.
We find that the \imsrgtwo{} gives a very similar prediction to the CCSD result of Ref.~\cite{Hagen2016NP_Ca48Skin}.
The \imsrgthreenseven{} prediction converges very quickly in terms of its model-space parameters, with essentially no difference between the $\emaxthreeb = 5$ and 6 predictions.
The resulting 0.005~fm correction to the neutron skin is about 3--4\,\% on the total neutron skin.

We find that the \imsrgthreenseven{} gives only small corrections to the \imsrgtwo{} for ground-state properties, solidifying the many \imsrgtwo{} studies of energies and charge radii where many-body method uncertainties have so far been unquantified.
We emphasize, however, that for precise predictions of small quantities or uncertainty quantification in exotic systems where no comparison to experiment is possible, the \imsrgthreenseven{} provides a systematic way to probe the many-body uncertainty of the \imsrgtwo{} truncation.

In the following, we consider one example of the \imsrgtwo{} truncation uncertainty in the difference between single-reference \imsrg{} and valence-space \imsrg{} calculations of the same system.
Closed (sub-)shell systems like \elem{Ca}{48} can be computed using the single-reference \imsrg{} and also using the valence-space \imsrg{}.
The two approaches employ the same reference state but differ in their decoupling conditions when solving the IMSRG equations.
Additionally, the \vsimsrg{} solves a part of the many-body problem exactly through the valence-space diagonalization.
In the limit of no many-body truncation the single-reference and valence-space \imsrg{} approaches yield identical results because both compute unitary transformations of the Hamiltonian that leave the eigenstates and eigenvalues unchanged.
Truncations of the \imsrg{} and \vsimsrg{},
such as, for example, the \imsrgtwo{} and \vsimsrgtwo{},
cause the unitary transformations to be approximate, and thus results differ for the two approaches.
This means that the predictions by the two methods for the same system will be slightly different, a result of the normal-ordered two-body truncation of the \imsrg{}.

This can be seen in Fig.~\ref{fig:VSIMSRGComparisonEnergy}, where the \imsrgtwo{} (gray, dot-dashed line in left panel) predicts a ground-state energy of $-415.8~\MeV$ while the \vsimsrgtwo{} (gray, dot-dashed line in center panel) predicts a ground-state energy of $-417.1~\MeV$.
The difference shown in the right panel is a result of the normal-ordered two-body approximation in both methods and the different decoupling conditions, generally reflective of the many-body truncation uncertainty.
Extending both methods to the normal-ordered three-body truncation via the \imsrgthreenseven{} and \vsimsrgthreenseven{} is expected to reduce the many-body truncation uncertainty and thus decrease the difference between the two methods.
This is exactly what one finds, as the difference comes down from 1.3 to 0.7~\MeV{} at $\emaxthreeb = 6$, $\ethreemax = 18$.

\begin{figure*}
    \includegraphics{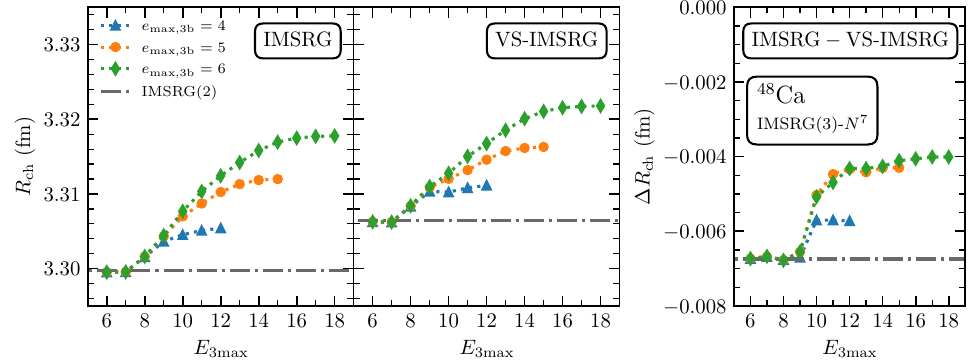}
    \caption{
        \label{fig:VSIMSRGComparisonRadius}
        The charge radius of \elem{Ca}{48} computed by the \imsrg{} and \vsimsrg{}.
        The left and center panels show \imsrgthreenseven{} results for increasing three-body truncations \emaxthreeb{} and \ethreemax{} obtained using the single-reference and valence-space \imsrg{} formulations, respectively, with the difference between the two results ($\Delta \rch = R_\text{ch,\imsrg} - R_\text{ch,\vsimsrg}$) displayed in the right panel.
    }
\end{figure*}

We observe similar behavior for the charge radius of \elem{Ca}{48} predicted by the \imsrg{} and \vsimsrg{} in Fig.~\ref{fig:VSIMSRGComparisonRadius}.
The \imsrgtwo{} predicts $\rch = 3.300~\fm$ while the \vsimsrgtwo{} predicts $\rch = 3.307~\fm$,
which differ by $\Delta \rch = R_\text{ch,\imsrg} - R_\text{ch,\vsimsrg} = -0.007~\fm$.
Going to the \imsrgthreenseven{} and \vsimsrgthreenseven{} truncations systematically reduces this difference,
yielding a difference of $\Delta \rch = -0.004~\fm$ for $\emaxthreeb = 6$, $\ethreemax=18$.

Overall, the reduced differences between \imsrg{} and \vsimsrg{} results when going from the (VS-)\imsrgtwo{} to the (VS-)\imsrgthreenseven{} indicate that, as expected, the many-body truncation uncertainties are being reduced.
We note here that for our \vsimsrgthreenseven{} we employ the approximation that valence-space three-body operators are truncated, motivated by the expected cluster hierarchy in ab initio calculations~\cite{Friman2011Book_FLT3N,Stroberg2019ARNPS_VSIMSRG}.
This approximation only affects the \vsimsrgthreenseven{} calculations, not the \imsrgthreenseven{} calculations,
and the fact that we find improved consistency between \vsimsrgthreenseven{} and \imsrgthreenseven{} results indicates that the effect of this approximation is small and under control in the cases we consider.
This is an important result because the inclusion of three-body operators in large-scale diagonalizations increases the cost by one to two orders of magnitude~\cite{Johnson2018_BIGSTICK}, which may be prohibitive in applications involving large valence spaces.
Nonetheless, it is also an important task for future work to test this approximation in tractable problems using available (no-core) shell-model solvers able to handle three-body interactions~\cite{Roth2009PRC_ITNCSM,Barrett2013PPNP_NCSMReview,Johnson2018_BIGSTICK}.

\begin{figure}
    \includegraphics{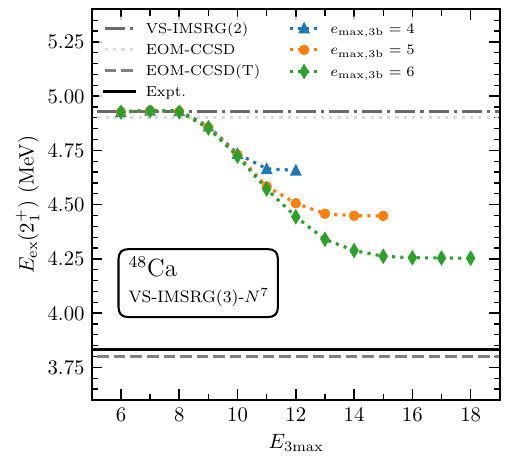}
    \caption{
        \label{fig:Ca48ExcitedState}
        The first $2^+$ excitation energy of \elem{Ca}{48} predicted by \vsimsrg{} calculations.
        The \vsimsrgtwo{} prediction (dot-dashed line) is compared to \vsimsrgthreenseven{} predictions for increasing \emaxthreeb{} and \ethreemax{},
        coupled-cluster values at the singles and doubles level (EOM-CCSD, dotted line)
        and including perturbative triples [EOM-CCSD(T), dashed line] from Ref.~\cite{Hagen2016PRL_Ni78},
        and the experimental value (solid line)~\cite{ENSDF2024Web}.
    }
\end{figure}

\begin{figure*}
    \includegraphics{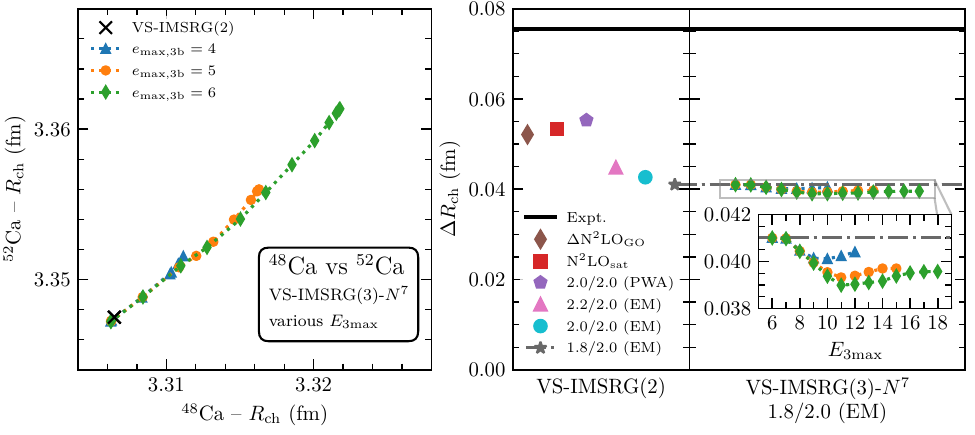}
    \caption{
        \label{fig:CaRadiiCorrelations}
        Comparison of charge radii of \elem{Ca}{48} and \elem{Ca}{52} for \vsimsrgtwo{} and \vsimsrgthreenseven{} calculations.
        \vsimsrgthreenseven{} predictions are given as a function of \emaxthreeb{} and \ethreemax{}.
        In the left panel we show predictions for both systems at the same truncation level.
        In the center and right panels, we consider the difference $\Delta R_\text{ch} = R_\text{ch}(\elem{Ca}{52}) - R_\text{ch}(\elem{Ca}{48})$.
        We show \vsimsrgtwo{} predictions for several Hamiltonians from chiral EFT in comparison to the much larger measured charge radius difference~\cite{GarciaRuiz2016NP_CaRadii} in the center
        and \vsimsrgthreenseven{} predictions for the \magic{} Hamiltonian on the right.
    }
\end{figure*}

A long-standing challenge for the \vsimsrgtwo{} has been the overprediction of $2^+$ excitation energies for closed-shell systems, notably \elem{Ca}{48}~\cite{Hagen2016PRL_Ni78,Simonis2017PRC_ChiralSaturationNuclei} and \elem{Ni}{78}~\cite{Hagen2016PRL_Ni78,Taniuchi2019N_Ni78,Tichai2023PLB_IMDMRG}.
In coupled-cluster theory, it has been established that in these cases the corrections due to triples are substantial~\cite{Hagen2016PRL_Ni78,Morris2018PRL_Sn100}.
In Fig.~\ref{fig:Ca48ExcitedState}, we revisit the $2^+$ energy of \elem{Ca}{48} with the \vsimsrgthreenseven{}.
We see that both CCSD [specifically equation-of-motion CCSD (EOM-CCSD)] and the \vsimsrgtwo{} substantially overpredict the experimental $2^+$ energy at $E_\text{ex}(2^+_1) = 3.832~\MeV$.
This is unusual for the \magic{} Hamiltonian, which generally accurately predicts spectra in \vsimsrgtwo{} calculations~\cite{Simonis2017PRC_ChiralSaturationNuclei, Stroberg2019ARNPS_VSIMSRG}.
Our \vsimsrgthreenseven{} calculations show that with increasing three-body model-space truncations \emaxthreeb{} and \ethreemax{} the $2^+$ energy comes down substantially.
At $\emaxthreeb = 6$, $\ethreemax = 18$, the $2^+$ is still far from fully converged, and a quantitative assignment of the \vsimsrgthreenseven{} prediction is not possible.
Nonetheless, the considerable \vsimsrgthreenseven{} corrections bring the $2^+$ down considerably into better agreement with coupled-cluster with triples [EOM-CCSD(T)] and experiment,
providing a substantially improved description of a key observable related to the closed-shell structure of \elem{Ca}{48}.
In these cases, it is clear that \vsimsrgtwo{} predictions have large many-body uncertainties and the \vsimsrgthreenseven{} is necessary for a precise description of spectra.

\subsection{Impact on charge radii}

\begin{figure*}
    \includegraphics{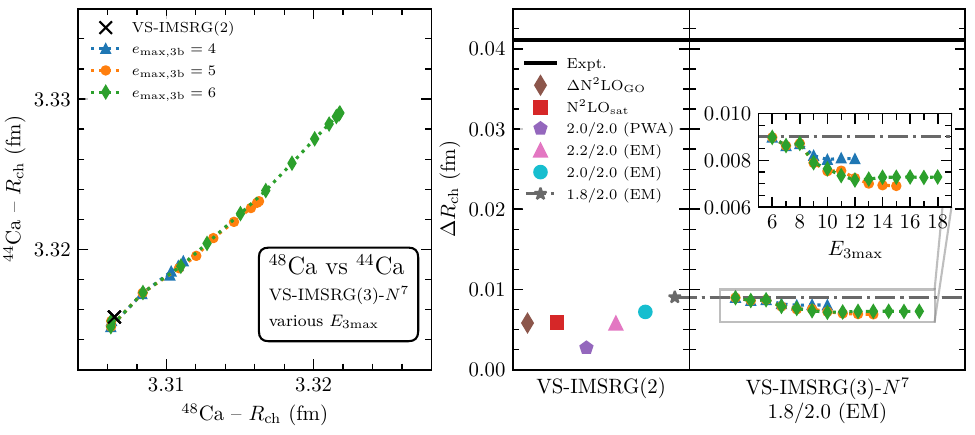}
    \caption{
        \label{fig:CaRadiiCorrelationsCa44}
        Same as Fig.~\ref{fig:CaRadiiCorrelations} but comparing \elem{Ca}{48} with \elem{Ca}{44} and showing the difference $\Delta R_\text{ch} = R_\text{ch}(\elem{Ca}{44}) - R_\text{ch}(\elem{Ca}{48})$.
    }
\end{figure*}

In the calcium isotopic chain, ab initio calculations are currently unable to predict most of the essential features of measured charge radii.
The charge radii of \elem{Ca}{40} and \elem{Ca}{48} are nearly identical, a feature that many ab initio calculations also reproduce~\cite{GarciaRuiz2016NP_CaRadii,Simonis2017PRC_ChiralSaturationNuclei,Hoppe2019PRC_ChiralMedMass,Miyagi2020PRC_MultiShellIMSRG}.
Those of \elem{Ca}{42}, \elem{Ca}{44}, and \elem{Ca}{46} are all considerably larger than either \elem{Ca}{40} or \elem{Ca}{48}, a feature unexplained by \vsimsrgtwo{} calculations~\cite{Miyagi2020PRC_MultiShellIMSRG} and so far unexplored by other many-body methods.
In the past, this has been phenomenologically explained as an effect of cross-shell excitations in the shell model with effective charges~\cite{Caurier2001PLB_ShellModelCalcium4048} or alternatively of particular pairing interactions in the context of energy density functional theory~\cite{Reinhard2017PRC_Fayans,Miller2019NP_SuperfluidCalcium}.

Another feature underpredicted by ab initio calculations is the surprisingly large charge radius of \elem{Ca}{52} relative to \elem{Ca}{48}~\cite{GarciaRuiz2016NP_CaRadii}, which puts into question the assignment of $N=32$ as a magic number in calcium isotopes.
Both in coupled-cluster theory at the CCSD (and triples) level and in the \vsimsrgtwo{}, for a broad range of chiral EFT Hamiltonians the charge radius of \elem{Ca}{52} relative to \elem{Ca}{48} is underpredicted by 33--50\,\%.
We revisit both of these questions with the \vsimsrgthreenseven{}, looking to gain insight into the effects of many-body corrections on this open puzzle.
We employ the \magic{} Hamiltonian, which notably considerably underpredicts absolute charge radii of medium-mass nuclei.
However, in the charge radius difference for two systems $\Delta \rch$, this systematic deficiency largely cancels and reproduction of charge radius differences or isotope shifts is once again much better.

In Fig.~\ref{fig:CaRadiiCorrelations}, we consider the charge radii of \elem{Ca}{48} and \elem{Ca}{52}.
In the left panel, we see predictions for the absolute charge radii, with the \vsimsrgtwo{} prediction indicated in the lower-left corner by the black cross.
The \vsimsrgthreenseven{} predictions for both systems for $\emaxthreeb = 4$, 5, 6 and $\ethreemax$ up to $3 \emaxthreeb$ are indicated as well.
We see that in both systems the \vsimsrgthreenseven{} gives corrections leading to larger charge radii,
and these corrections are very similar in both systems.
This means that the many-body and three-body model-space uncertainties in both systems are highly correlated and, as a result, cancel when we consider the difference.
We see this feature in the right panel of Fig.~\ref{fig:CaRadiiCorrelations},
where the \vsimsrgtwo{} result (the gray, dot-dashed line) underpredicts the experimental value (the black line) by nearly 50\,\%.
On the scale shown, the \vsimsrgthreenseven{} corrections to the difference are extremely small,
and looking at the inset we see that the difference is essentially converged in model-space size at $\emaxthreeb= 6$, $\ethreemax = 18$,
and the corrections change the charge radius difference by less than 10\,\%, notably towards a smaller, not larger difference.

If we compare this to the chiral EFT uncertainty, represented in the center panel by \vsimsrgtwo{} predictions using several well-established Hamiltonians~\cite{Hebeler2011PRC_SRG3NFits,Ekstrom2015PRC_N2LOsat,Jiang2020PRC_DN2LOGO},
the many-body uncertainty of $\approx5\,\%$ on the experimental value is much smaller than the nearly 25\,\% due to variation of the Hamiltonian employed.
Our results indicate that a resolution of the theoretical underprediction of the relatively large charge radius of \elem{Ca}{52} is not offered by the \vsimsrgthreenseven{}.
This further motivates the open question of how to improve or adjust nuclear forces in chiral EFT to reproduce this large radius difference.
On the other hand, it is also still possible that considerable corrections are missed by the \vsimsrgthreenseven{} truncation and require going to higher orders in the many-body expansion.

We find a similar picture for the charge radius of \elem{Ca}{44} relative to that of \elem{Ca}{48} in Fig.~\ref{fig:CaRadiiCorrelationsCa44}.
In the left panel, we see that the many-body corrections to the charge radii of both systems are again very strongly correlated, meaning that they cancel in the difference shown in the right panel.
There we see that the experimental charge radius difference $\Delta \rch = 0.0411~\fm$ is vastly underpredicted by the \vsimsrgtwo{}, predicting a difference $\Delta \rch = 0.009~\fm$.
The \vsimsrgthreenseven{} corrections are small and converge quickly with model-space size giving a difference $\Delta \rch = 0.0073~\fm$ at $\emaxthreeb = 6$, $\ethreemax = 18$.
This correction is once again much smaller than the effect of Hamiltonian variation as shown in the center panel,
where \vsimsrgtwo{} predictions with different Hamiltonians vary by about 20\,\% on the experimental value.

In this case, one effect we do not investigate is the effect of opening up the valence space to allow for cross-shell excitations as was done in shell-model studies~\cite{Caurier2001PLB_ShellModelCalcium4048}.
This was investigated in the \vsimsrgtwo{} in Ref.~\cite{Miyagi2020PRC_MultiShellIMSRG} and led to no appreciable change in the trend of charge radii between \elem{Ca}{40} and \elem{Ca}{48}.
It is possible that the \vsimsrgthreenseven{} with a multishell valence space will change this picture significantly.
Again, we find that \vsimsrgthreenseven{} many-body corrections on the radius difference are small and unable to explain current discrepancies with data.
This motivates the developments of further improvements to the many-body method and improved Hamiltonians.

\subsection{Improved excitation spectra}

\begin{figure*}
    \includegraphics{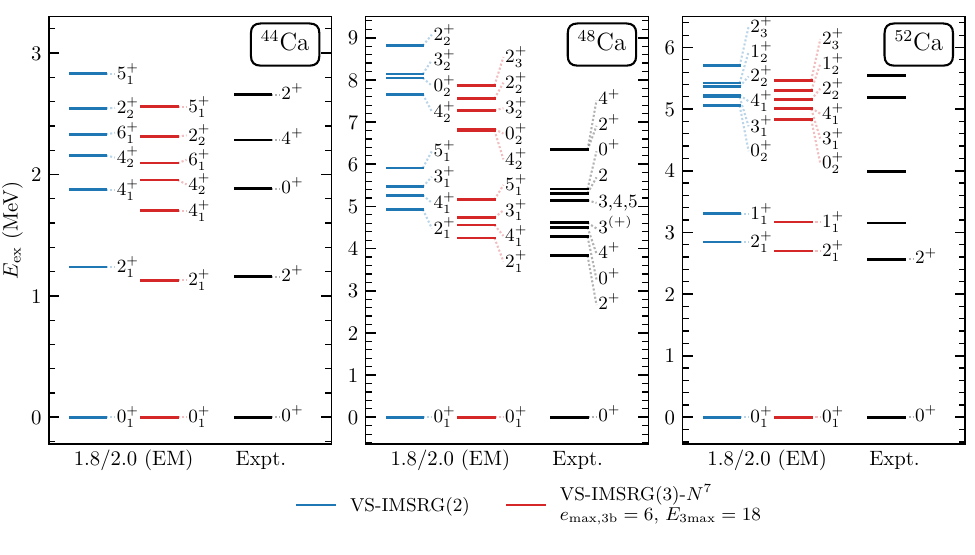}
    \caption{
        \label{fig:CaSpectra}
        Low-lying excitation spectra of positive-parity states of \elem{Ca}{44} (left), \elem{Ca}{48} (center), and \elem{Ca}{52} (right).
        \vsimsrgtwo{} results are compared with \vsimsrgthreenseven{} predictions using the truncations $\emaxthreeb = 6$, $\ethreemax = 18$ and experiment~\cite{ENSDF2024Web}.
        In \elem{Ca}{52} we show all states as most states do not yet have a spin and parity assignment.
    }
\end{figure*}

In Fig.~\ref{fig:CaSpectra}, we compare the spectrum of positive-parity states predicted by the \vsimsrgtwo{} and \vsimsrgthreenseven{} (using our largest model-space truncation $\emaxthreeb = 6$, $\ethreemax = 18$) for \elem{Ca}{44}, \elem{Ca}{48}, and \elem{Ca}{52}.
We see a similar behavior in all three systems,
namely that the \vsimsrgthreenseven{} brings all states down in the spectrum, essentially reducing their energies by a common factor.
This reduction factor appears to be different in all three systems, largest in \elem{Ca}{48} and smallest in \elem{Ca}{52}.
In all systems, the \vsimsrgthreenseven{} predictions for the first $2^+$ energy approach the experimental value.
This systematic trend is also seen for many other states, such as the lowest $4^+$, $3^+$, and $5^+$ states in \elem{Ca}{48} and the experimentally unassigned $1^+$ state in \elem{Ca}{52}.
It is possible to understand this trend as a relative increase in the ground-state energy in the spectrum due to reduced valence-space matrix elements coupling particle and hole states as was observed in Ref.~\cite{Stroberg2024PRC_IMSRG3}.
We see that such off-diagonal matrix elements are also generally smaller in our \vsimsrgthreenseven{} calculations than in our \vsimsrgtwo{} calculations.

A low-lying state that we do not reproduce in \elem{Ca}{48} is the first excited $0^+$ state.
The reproduction of low-lying $0^+$ states in closed-shell light and medium-mass nuclei is a long-standing problem (see, e.g., Refs.~\cite{Brown1966NP_O16, Federman1969PR_Ca40, Sakakura1976PLB_Ca40} for discussions on \elem{O}{16} and \elem{Ca}{40}).
The phenomenology to explain these states is mixed, relying on excitations into deformed states or complicated many-particle excitations.
Regardless, even at the \vsimsrgthreenseven{} level, we do not capture whatever physics lies behind the lowest excited $0^+$ state.
Our first $0^+$ excited state (which may not be the intruder state with more complex many-body configurations) lies far above the experimental energies for both the first and second $0^+$ excited states.
However, the large \vsimsrgthreenseven{} corrections suggest that additional large many-body corrections from the \vsimsrgthree{} or beyond may play an important role here.

\subsection{Implications for many-body uncertainties}

\begin{table}
    \caption{
        \label{tab:IMSRG3Corrections}
        Comparison of \imsrgtwo{} and \imsrgthreenseven{} predictions for several observables in several calcium isotopes.
        We consider both single-reference \imsrg{} calculations (top) and valence-space \imsrg{} calculations (bottom), showing the \imsrgtwo{} result, the \imsrgthreenseven{} correction (at $\emaxthreeb = 6$, $\ethreemax= 18$), and the percentage change induced by the \imsrgthreenseven{} correction.
        \imsrgtwo{} correlation energies $\ecorr = E_\text{IMSRG(2)} - E_\hf$ and excitation energies are given in MeV.
        Charge radii and neutron skins are given in fm.
    }
    \begin{ruledtabular}
    \begin{tabular}{lrrrr}
         \textbf{IMSRG} & \imsrgtwo{} & $\Delta$\imsrgthreenseven{} & \%  \\ 
         \\[-2.5ex] \hline & \\[-2.5ex]
         \ecorr(\elem{Ca}{40}) & $-96.9$ & $-1.7$ & 1.7 \\
         \ecorr(\elem{Ca}{48}) & $-112.2$ & $-1.8$ & 1.6 \\
         \ecorr(\elem{Ca}{52}) & $-119.9$ & $-2.0$ & 1.6 \\
         \\[-2.5ex] \hline & \\[-2.5ex]
         \rch(\elem{Ca}{40}) & $3.319$ & 0.011 & 0.3 \\
         \rch(\elem{Ca}{48}) & $3.300$ & 0.018 & 0.5 \\
         \rch(\elem{Ca}{52}) & $3.340$ & 0.017 & 0.5 \\
         \\[-2.5ex] \hline & \\[-2.5ex]
         \rskin(\elem{Ca}{40}) & $-0.041$ & $-0.001$ & 2.5\\
         \rskin(\elem{Ca}{48}) & $0.145$ & 0.005 & 3.1 \\
         \rskin(\elem{Ca}{52}) & $0.283$ & 0.003 & 1.2\\
         \\[-2.5ex]\hline\hline\\[-2.5ex]
         \textbf{VS-IMSRG} & \vsimsrgtwo{} & $\Delta$\vsimsrgthreenseven{} & \% \\ 
         \\[-2.5ex] \hline & \\[-2.5ex]
         \ecorr(\elem{Ca}{44}) & $-108.2$ & $-1.4$ & 1.3 \\
         \ecorr(\elem{Ca}{48}) & $-113.5$ & $-1.2$ & 1.1 \\
         \ecorr(\elem{Ca}{52}) & $-121.4$ & $-1.3$ & 1.1 \\
         \\[-2.5ex] \hline & \\[-2.5ex]
         \rch(\elem{Ca}{44}) & $3.316$ & 0.013 & 0.4 \\
         \rch(\elem{Ca}{48}) & $3.306$ & 0.015 & 0.5 \\
         \rch(\elem{Ca}{52}) & $3.347$ & 0.014 & 0.4\\
         \\[-2.5ex] \hline & \\[-2.5ex]
         \rskin(\elem{Ca}{44}) & $0.070$ & $0.005$ & 7.4\\
         \rskin(\elem{Ca}{48}) & $0.142$ & 0.007 & 5.2 \\
         \rskin(\elem{Ca}{52}) & $0.278$ & 0.008 & 2.8\\
         \\[-2.5ex] \hline & \\[-2.5ex]
         \elem{Ca}{44} -- $E_\text{ex}(2^+_1)$ & $1.238$ & $-0.110$ & $-8.9$\\
         \elem{Ca}{44} -- $E_\text{ex}(4^+_1)$ & $1.875$ & $-0.172$ & $-9.2$\\
         \elem{Ca}{44} -- $E_\text{ex}(4^+_2)$ & $2.156$ & $-0.201$ & $-9.3$\\
         \elem{Ca}{48} -- $E_\text{ex}(2^+_1)$ & $4.930$ & $-0.677$ & $-13.7$\\
         \elem{Ca}{48} -- $E_\text{ex}(0^+_2)$ & $8.044$ & $-1.211$ & $-15.0$\\
         \elem{Ca}{48} -- $E_\text{ex}(4^+_1)$ & $5.266$ & $-0.704$ & $-13.4$\\
         \elem{Ca}{52} -- $E_\text{ex}(2^+_1)$ & $2.844$ & $-0.148$ & $-5.2$\\
         \elem{Ca}{52} -- $E_\text{ex}(1^+_1)$ & $3.302$ & $-0.133$ & $-4.0$\\
         \elem{Ca}{52} -- $E_\text{ex}(0^+_2)$ & $5.055$ & $-0.225$ & $-4.5$
    \end{tabular}
    \end{ruledtabular}
\end{table}

Our results show that in many cases, especially for ground-state properties, the \imsrgthree{} corrections we compute are small and not essential for a quantitative description of the system.
Nonetheless, it is still important to quantify these uncertainties, and here we shed light on the approximate order of magnitude of \imsrgthree{} corrections in medium-mass nuclei.
In coupled-cluster theory, there is the well-established rule of thumb that CCSD captures 90\,\% of the correlation energy $E_\text{exact} - E_\hf{}$, triples account for an additional 9\,\%, and the rest comes from high-order effects~\cite{Hagen2014RPP_CCReview}.
Additionally, CC calculations typically estimate the effect of triples for charge radii to be on the order of 1\,\%~\cite{Hagen2016NP_Ca48Skin,Simonis2019EPJA_CCEMResponses}.
These uncertainty estimates are rough, but generally applicable owing to the fact that coupled-cluster is a size-extensive method.
As the \imsrg{} is also size extensive, the insights we provide here will be more broadly applicable but should still be considered a rough rule of thumb that can be refined by actually performing a \vsimsrgthreenseven{} calculation.

Table~\ref{tab:IMSRG3Corrections} lists \imsrg{} and \vsimsrg{} results for several quantities computed with the (VS-)\imsrgtwo{} and (VS-)\imsrgthreenseven{} in \elem{Ca}{40}, \elem{Ca}{44}, \elem{Ca}{48}, and \elem{Ca}{52}.
Our \imsrgthreenseven{} corrections are always computed using our largest model-space truncation,
$\emaxthreeb = 6$, $\ethreemax = 18$.
For \imsrgtwo{} correlation energies $\ecorr = E_\text{\imsrgtwo} - E_\hf$, we find small corrections from the \imsrgthreenseven{} of around $2~\MeV$, which correspond to 1--2\,\% corrections to the correlation energy.
Recall that our remaining model-space uncertainty is estimated to be $1.5~\MeV$, meaning that this percentage is likely in the range 2--3\,\% for fully converged calculations.
We note that the \vsimsrgthreenseven{} corrections are generally smaller than the \imsrgthreenseven{} corrections.
We understand this to be a result of the exact treatment of part of the many-body problem in the valence-space diagonalization.
In \elem{Ca}{48}, the valence-space diagonalization accounts for around $-78~\MeV$ of the binding energy of the system.
At the same time, the valence-space decoupling is more complicated than the single-reference decoupling, leading to larger missing induced three-body interactions, which the \vsimsrgthreenseven{} partially resolves.
This situation seems to balance out such that \imsrgthreenseven{} corrections to the correlation energy are smaller in valence-space calculations than in single-reference calculations.
Our estimate for the general \vsimsrgtwo{} uncertainty on the correlation energy is thus 1--2\,\%.

Charge radii are quantitatively well described at the mean-field level, with only small corrections from the \imsrgtwo{}.
The \imsrgthreenseven{} corrections are also small (although not fully converged at $\emaxthreeb = 6$, $\ethreemax = 18$).
We see that on the total charge radius, the \imsrgthreenseven{} and \vsimsrgthreenseven{} provide corrections of around 0.5\,\% at our truncations.
Accounting for a similar further increase from reaching full model-space convergence, we estimate the \imsrgtwo{} and \vsimsrgtwo{} uncertainty for charge radii to be 1--1.5\,\%.
For the neutron skin, we find larger relative corrections, which is to be expected as \rskin{} is a differential quantity that is relatively small.
We already benefit from significant cancellations between correlated changes to the point-proton and point-neutron radii to give a smaller (VS-)\imsrgthreenseven{} correction to \rskin{} than for instance \rch.
It is likely based on Fig.~\ref{fig:IMSRG3GroundState} that our (VS-)\imsrgthreenseven{} predictions for neutron skins are nearly fully converged at $\emaxthreeb = 6$, $\ethreemax = 18$,
so we conservatively estimate a (VS-)\imsrgtwo{} uncertainty of 5--7.5\,\% on neutron skins.

Our work establishes that the \vsimsrgthreenseven{} brings in important corrections necessary for a quantitative description of the $2^+$ energy of \elem{Ca}{48} and similar corrections for many other excited states.
In Table~\ref{tab:IMSRG3Corrections}, we see that the overall reduction of the spectrum by the \vsimsrgthreenseven{} for each system is visible in the percentages given on the right.
In \elem{Ca}{44}, the energies of states are consistently reduced by around 9\,\%.
In \elem{Ca}{48} this effect is larger, around 13.5--15\,\%, and in \elem{Ca}{52} this effect is smaller, only around 4--5\,\%.
It is somewhat surprising and interesting that all states are modified similarly,
which may be connected back to leading \imsrgthree{} contributions being related to modified single-particle energies~\cite{He2024PRC_IMSRG3f2}.
Nonetheless, it is clear that the \vsimsrgthree{} gives important (but probably not larger than 25\,\%) corrections to excitation energies.
The actual size of these corrections is not size extensive and system dependent, so our estimates here are not easily transferable to other systems.

\section{Conclusion}
\label{sec:Conclusion}

We study the structure of calcium isotopes using the \imsrgthreenseven{} and \vsimsrgthreenseven{}, which provide more precise solutions to the many-body Schrödinger equation than the \imsrgtwo{} and \vsimsrgtwo{}, respectively.
We find that this improved precision gives small corrections for ground-state properties that are very consistent with benchmarks from coupled-cluster theory including triples.
It also improves the consistency between the \imsrg{} and \vsimsrg{} approaches, which differ slightly due to the many-body truncation employed, systematically reducing difference between the two approaches.
When we turn to the $2^+$ energy of \elem{Ca}{48}, we find that the \vsimsrgthreenseven{} provides large corrections that bring the $2^+$ excitation energy down into much better agreement with experiment and also coupled-cluster with triples, improving the description of the shell closure at $N=28$.

We find that \vsimsrgthreenseven{} corrections to charge radii in \elem{Ca}{44}, \elem{Ca}{48}, and \elem{Ca}{52} are strongly correlated.
This results in only very small changes to the charge radius differences between the systems, much smaller than the chiral EFT uncertainty explored by using different Hamiltonians.
This indicates that the \imsrgthreenseven{} approximation does not resolve existing theoretical challenges in describing charge radius trends.
These may instead be due to systematic deficiencies in currently used chiral EFT Hamiltonians
or alternatively due to many-body effects not captured by the \imsrgthreenseven{}, motivating the development of further improvements to many-body methods.

Based on the size extensivity of the \imsrg{}, we are able to provide general estimates for many-body uncertainties at the \imsrgtwo{} level based on the \imsrgthreenseven{} corrections we compute in several systems. For soft Hamiltonians, we estimate the \imsrgtwo{} has a 2--3\,\% uncertainty on the correlation energy, a 1--1.5\,\% uncertainty on the charge radius (and also point-neutron radius), and a 5--7.5\,\% uncertainty on the neutron skin.
We find that the \vsimsrgthreenseven{} systematically lowers all excitation energies in the spectrum, but by varying amounts in different systems, preventing a general uncertainty estimate.

This work establishes the \imsrgthreenseven{} to explore many-body uncertainties and to improve on \imsrgtwo{} predictions for ground-state and excited-state properties.
Convergence in medium-mass nuclei is challenging, and the extension to heavier systems will require innovative computational approaches~\cite{Hagen2012PTPS_CCDistribution} and more effective model-space truncations than the \emaxthreeb{}, \ethreemax{} truncations employed in this work~\cite{Novario2020PRC_DeformedCC}.
Recently developed factorized approximations to the \imsrgthree{} offer a complementary way to explore many-body uncertainties~\cite{He2024PRC_IMSRG3f2}, both by cheaply approximating the \imsrgthree{} and via appropriate extensions possibly capturing leading \imsrgfour{} effects.
Exploring all of these approaches will be important to making high-precision \imsrg{} calculations more routine, which is also a key step towards statistically robust many-body uncertainty quantification.

\begin{acknowledgments}

We thank Gaute Hagen, Jan Hoppe, Gustav Jansen, and Thomas Papenbrock for helpful discussions, Francesca Bonaiti for providing additional coupled-cluster values for \elem{Ca}{48}~\cite{Bonaiti2024private}, and Martin Hoferichter and Frederic Noël for pointing out the error in Ref.~\cite{Ong2010PRC_SpinOrbit} and benchmarking the correction described in Appendix~\ref{app:RadiusOps}.
M.H.~gratefully acknowledges the hospitality of the Argonne National Laboratory nuclear theory group in 2022.
This work was supported in part by the European Research Council (ERC) under the European Union's Horizon 2020 research and innovation programme (Grant Agreement No.~101020842),
by the Deutsche Forschungsgemeinschaft (DFG, German Research Foundation) -- Projektnummer 279384907 -- SFB 1245,
by the U.S.\ Department of Energy, Office of Science, Office of Advanced Scientific Computing Research and Office of Nuclear Physics, Scientific Discovery through Advanced Computing (SciDAC) program (SciDAC-5 NUCLEI), 
by the Laboratory Directed Research and Development Program of Oak Ridge National Laboratory, managed by UT-Battelle, LLC, for the U.S.\ Department of Energy, by JST ERATO Grant No.~JPMJER2304, Japan, and by the U.S.\ National Science Foundation Grant No PHY-2340834-01.
This research used resources of the Oak Ridge Leadership Computing Facility located at Oak Ridge National Laboratory, which is supported by the Office of Science of the Department of Energy under contract No.~DE-AC05-00OR22725.
The authors gratefully acknowledge the Gauss Centre for Supercomputing e.V.~\cite{GaussCentre} for funding this project by providing computing time through the John von Neumann Institute for Computing (NIC) on the GCS Supercomputer JUWELS at Jülich Supercomputing Centre (JSC) and the computing time provided to them on the high-performance computer Lichtenberg II at TU Darmstadt, funded by the German Federal Ministry of Education and Research (BMBF) and the State of Hesse.
This work has been partially supported by U.S.~DOE Grant No.~DE-FG02-13ER41967.
ORNL is managed by UT-Battelle, LLC, under Contract No.~DE-AC05-00OR22725 for the U.S.~Department of Energy (DOE).
The US government retains and the publisher, by accepting the article for publication, acknowledges that the US government retains a nonexclusive, paid-up, irrevocable, worldwide license to publish or reproduce the published form of this manuscript, or allow others to do so, for US government purposes. DOE will provide public access to these results of federally sponsored research in accordance with the DOE Public Access Plan~\cite{DOEPublicAccessPlan}.

\end{acknowledgments}

\appendix

\section{Hybrid Hartree-Fock+NAT basis}
\label{app:HFNAT}

The standard basis choice for nuclear structure calculations is the Hartree-Fock basis:
\begin{equation}
    \ket{p}_\hf = \ket{n l j m_t}_\hf = \sum_{n'} C^{\hf, l j m_t}_{n n'} \ket{n' l j m_t}_\ho,
\end{equation}
constructed from \ho{} states $\ket{p}_\ho = \ket{n l j m_t}_\ho$.
We employ an angular momentum, parity, and isospin conserving scheme, allowing us to ignore the trivial dependence on $m_j$ and to restrict the \ho{} to \hf{} mixing to only the principal quantum numbers $n$, $n'$.
A successful alternative is the \nat{} basis~\cite{Tichai2019PRC_NAT}:
\begin{equation}
    \ket{p}_\nat = \ket{n l j m_t}_\nat = \sum_{n'} C^{\nat, l j m_t}_{n n'} \ket{n' l j m_t}_\ho.
\end{equation}
The \nat{} basis has been very successful in accelerating convergence in nuclear structure calculations~\cite{Tichai2019PRC_NAT, Novario2020PRC_DeformedCC, Hoppe2021PRC_NAT}.
At the same time, it naively requires one to give up the canonical \hf{} reference state to produce a single Slater-determinant reference state in the new basis,
which can lead to unexpected changes in predicted energies~\cite{Hoppe2021PRC_NAT}.

One way around this is to simply combine the two bases as we do in this work.
We call this the hybrid HF+NAT basis.
We start from the \hf{} basis $\ket{p}_\hf$ and the \nat{} basis $\ket{p}_\nat$ from second-order many-body perturbation theory, where for the \nat{} basis we have sorted our basis such that states with the highest magnitude of \nat{} occupation number (see Ref.~\cite{Hoppe2021PRC_NAT} for details) are assigned the lowest principal quantum number.
We then construct a basis systematically, starting from the occupied states in our reference state.
For these occupied states, we simply choose the \hf{} state:
\begin{equation}
    \ket{p} = \ket{p}_\hf.
\end{equation}
The remaining states are taken from the \nat{} basis, but we need to account for the fact that the lowest \hf{} states and the higher \nat{} states are not properly orthogonal.
We ensure this by performing a Gram-Schmidt orthogonalization of each \nat{} state with respect to all states already included in our basis.
In most reasonable cases the \hf{} and \nat{} occupied states are very similar, so the orthogonalization only changes the \nat{} states slightly and is merely a formality.
This preserves the beneficial convergence properties of the \nat{} basis while also allowing one to work with an \hf{} reference state.

\begin{figure}
    \centering
    \includegraphics{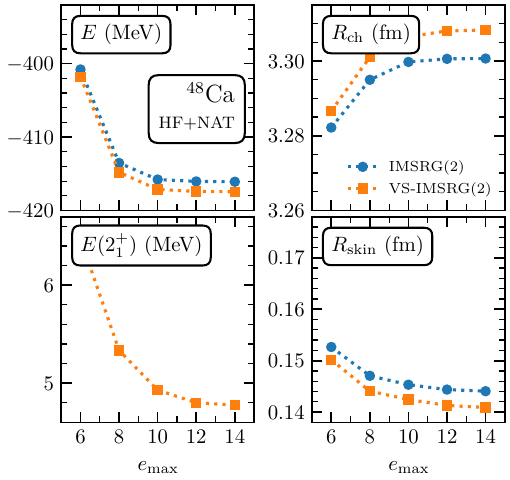}
    \caption{    
        \label{fig:Ca48IMSRG2}
        Model-space convergence of \imsrgtwo{} predictions of properties of \elem{Ca}{48} for various truncations of the computational basis \emax{}.
        The ground-state energy (top left), charge radius (top right), and neutron skin (bottom right) are computed using both the IMSRG(2) and VS-IMSRG(2), while the first $2^+$ excitation energy (bottom left) is only computed using the VS-IMSRG(2).
    }
\end{figure}

In the construction above, we start from an \ho{} basis consisting of 17 major shells ($\emaxmeanfield = 16$) with three-body matrix elements truncated at $\ethreemaxmeanfield = 24$.
We are left with an HF+NAT basis with an intrinsic truncation of $\emax=16$.
Realistically, however, following this basis optimization one can truncate the HF+NAT basis to much smaller \emax{} and still obtain converged results.
We see this in Fig.~\ref{fig:Ca48IMSRG2}, where we investigate the dependence of \imsrgtwo{} calculations on the truncation of the HF+NAT basis \emax{} for various quantities studied in this work.
We find that for our purposes $\emax=10$ is sufficiently converged.

\section{Radius operators and correction to spin-orbit charge radius contribution}
\label{app:RadiusOps}

In Eqs.~\eqref{eq:Rch} and~\eqref{eq:Rskin}, $\langle \rptwo \rangle$, $\langle \rntwo \rangle$, and $\langle \rsotwo \rangle$ are the nuclear point-proton radius, nuclear point-neutron radius, and nuclear spin-orbit radius expectation values, respectively.

$R^2_\text{p}$ has the operator expression
\begin{align}
  \label{eq:Rp2}
  R^2_\text{p} & =
  \sum_i \Big[
    (1 + \tau_i) \frac{1}{2Z}
    \Big( 1 - \frac{2}{A}\Big)
    +
    \frac{1}{A^2}
    \Big] \mathbf{r}_i^2 \nonumber \\ & \quad
  +\sum_{i<j} \Big[
    \frac{2}{A^2}
    - \frac{2}{AZ}
    \Big(1 + \frac{\tau_i + \tau_j}{2} \Big)
    \Big] \mathbf{r}_i \cdot \mathbf{r}_j\,,
\end{align}
with the number of protons, $Z$, and number of nucleons, $A$.
$\tau_i$ gives twice the isospin projection of the particle species in state $i$
\begin{equation}
  \tau_i = 
  \begin{cases}
    \hphantom{-}1 & \text{if $i$ is a proton}, \\
    -1 & \text{if $i$ is a neutron}.
  \end{cases}
\end{equation}
$r^2_\text{so}$ has the operator expression
\begin{equation}
  \label{eq:Rso2}
  r^2_\text{so} = - \sum_i \frac{\mu_i - Q_i / 2}{Z M^2} (\kappa_i + 1)\,,
\end{equation}
with the nucleon mass $M$.
$\mu_i$ gives the magnetic moment of the particle species in state $i$ (in units of the nuclear magneton $\mu_\text{N}$)
\begin{equation}
  \mu_i = 
  \begin{cases}
    \hphantom{-}2.793 & \text{if $i$ is a proton} \\
    -1.913 & \text{if $i$ is a neutron},
  \end{cases}
\end{equation}
$Q_i$ gives the charge of the particle species in state $i$ (in units of elementary charge $e$)
\begin{equation}
  Q_i = 
  \begin{cases}
    1 & \text{if $i$ is a proton} \\
    0 & \text{if $i$ is a neutron},
  \end{cases}
\end{equation}
and $\kappa_i$ gives the spin-orbit correction
\begin{equation}
  \kappa_i = 
  \begin{cases}
    l_i & \text{if $j_i = l_i - \frac{1}{2}$} \\
    -(l_i + 1) & \text{if $j_i = l_i + \frac{1}{2}$}.
  \end{cases}
\end{equation}
Our expression for $\rsotwo$ in Eq.~\eqref{eq:Rso2} corrects an error in the expression of Ref.~\cite{Ong2010PRC_SpinOrbit} pointed out by Martin Hoferichter.
The correction is simple:
\begin{equation*}
  \mu_i - Q_i \rightarrow \mu_i - Q_i / 2\,.
\end{equation*}
We verified the error in the previous expression and validity of the correction by calculations of the $\Phi''(q)$ nuclear responses, which are related to the spin-orbit radius at $q=0$~\cite{Fitzpatrick2013JCAP_NuclearResponsesDM, Hoferichter2020PRD_CEvNS, Hu2022PRL_DMResponses, Noel2024JHEP_ChargeDensities,Heinz2024inprep_MuToEResponses}.

$R^2_\text{n}$ has the operator expression
\begin{align}
  \label{eq:Rn2}
  R^2_\text{n} & =
  \sum_i \Big[
    (1 - \tau_i) \frac{1}{2N}
    \Big( 1 - \frac{2}{A}\Big)
    +
    \frac{1}{A^2}
    \Big] \mathbf{r}_i^2 \nonumber \\ & \quad
  +\sum_{i<j} \Big[
    \frac{2}{A^2}
    - \frac{2}{AN}
    \Big(1 - \frac{\tau_i + \tau_j}{2} \Big)
    \Big] \mathbf{r}_i \cdot \mathbf{r}_j\,,
\end{align}
with the number of neutrons, $N$.

\bibliography{ref}

\end{document}